\documentclass{article}
\usepackage[utf8]{inputenc}
\usepackage{authblk}
\usepackage{setspace}
\usepackage[margin=1.25in]{geometry}
\usepackage{graphicx}
\graphicspath{ {./figures/} }
\usepackage{subcaption}
\usepackage{amsmath}
\usepackage{lineno}

\usepackage[style=ieee, 
citestyle=numeric-comp,
sorting=none]{biblatex}
\title{Exploring the Potential of Integrated Optical Sensing and Communication (IOSAC) Systems with Si Waveguides for Future Networks}

\author[1$\dag$]{Xiangpeng Ou}
\author[2$\dag$]{Ying Qiu}
\author[2,3]{Ming Luo}
\author[1]{Fujun Sun}
\author[1]{Peng Zhang}
\author[1]{Gang Yang}
\author[1]{Junjie Li}
\author[1]{Jianfeng Gao}
\author[1]{Xiaobin He}
\author[1]{Anyan Du}
\author[1]{Bo Tang}
\author[1]{Bin Li}
\author[2]{Zichen Liu}
\author[1]{Zhihua Li}
\author[1]{Ling Xie}
\author[2,3]{Xi Xiao}
\author[1]{Jun Luo}
\author[1]{Wenwu Wang}
\author[2,3*]{Jin Tao}
\author[1*]{Yan Yang}

\affil[1]{Institute of Microelectronics, Chinese Academy of Sciences, Beijing,100029, China}

\affil[2]{State Key Laboratory of Optical Communication Technologies and Networks, and National Information Optoelectronics Innovation Center, China Information Communication Technologies Group Corporation (CICT), Wuhan, 430074,China}

\affil[3]{Peng Cheng Laboratory, Shenzhen, 518055,China}

\affil[*]{Corresponding author. Email: yyang10@ime.ac.cn}

\affil[$\dag$]{These authors contributed equally to this work.}

\date{}

\begin{document}

\maketitle

\begin{abstract}
Advanced silicon photonic technologies enable integrated optical sensing and communication (IOSAC) in real time for the emerging application requirements of simultaneous sensing and communication for next-generation networks. Here, we propose and demonstrate the IOSAC system on the silicon nitride (SiN) photonics platform. The IOSAC devices based on microring resonators are capable of monitoring the variation of analytes, transmitting the information to the terminal along with the modulated optical signal in real-time, and replacing bulk optics in high-precision and high-speed applications. By directly integrating SiN ring resonators with optical communication networks, simultaneous sensing and optical communication are demonstrated by an optical signal transmission experimental system using especially filtering amplified spontaneous emission spectra. The refractive index (RI) sensing ring with a sensitivity of 172 nm/RIU, a figure of merit (FOM) of 1220, and a detection limit (DL) of 
\(8.2 \times 10^{-6}\) RIU is demonstrated. Simultaneously, the 1.25 Gbps optical on-off-keying (OOK) signal is transmitted at the concentration of different NaCl solutions, which indicates the bit-error-ratio (BER) decreases with the increase in concentration. The novel IOSAC technology shows the potential to realize high-performance simultaneous biosensing and communication in real time and further accelerate the development of IoT and 6G networks. 
\end{abstract}


\section{Introduction}
With the explosive growth of Internet of Things (IoTs) which capable of sensing and communication (S\&C), the innova-tive characteristics of next-generation networks should be defined [1-4]. Currently, sensing and communication modules are individually accomplished with limited hardware intersection, rare mutual assistance and integration. This leads to situations where systems need to work with both communication and sensing, such as autonomous driving and drone management, requiring access to two different networks at the same time to meet practical needs. However, connecting two independent networks leads to high costs and unsynchronized information. The realization of the sensing function relies on the observation of noise, while the communication function focuses on identifying the encoded signal and then recovering from the reception of noise, but the two modules are based on similar hardware structures and signal processing modes. Therefore, S\&C systems could be jointly designed, optimized, and integrated to efficiently utilize congested re-sources and even pursue mutual benefits, which motivates a new technology of integrated sensing and communications (ISAC) [1, 5]. The exiting large volume communications infrastructures can be adopted to achieve ISAC technology with minimal standard modifications, and promote mutual benefits, which reduce both hardware and signaling costs as well as compact product size and enhanced spectral efficiency. 

Optics and photonic devices as promising candidates to achieve high-performance communication systems and high-sensitivity sensors, have been wide explored [6-9]. However, the previous work only focusses on individually optical communication and sensing. Hence, we propose the concept of integrated optical sensing and communication (IOSAC) namely simultaneous optical S\&C on one architecture based on optic and photonic technologies in this work. For the past decades, optical communication networks have played a dominant role in the high-bandwidth modern communication systems [10-13]. The sensing functionality as an increasingly essential factor embedded into ubiquitous traditional optical communication network in everywhere has been recognized as a key paradigm that enables a wide variety of applications, such as smart sensors, environmental monitoring, intelligent transportation and manufacturing, etc. [14-16]. 

Silicon photonics is promising to enable photonic integrated circuits (PICs) with high densities, cutting-edge func-tionality, and portability, following the path of electronic integrated circuits (EICs) [17-20]. The performance and integra-tion complexity of silicon photonic devices have been intensively developed and boosted by the compatible advanced manufacturing technologies of the standard complementary metal-oxide-semiconductor (CMOS) industry in the last decade [21-24]. It enables a wide range of emerging applications from data centers and telecommunication to on-chip sensors [25, 26]. Silicon photonics transceivers have been successfully commercialized and are playing an increasingly important role in optical communication networks, including optical fiber communication and wireless optical communication [27-30]. However, the sensing functionality based on silicon photonic technologies have not been widely used in the communica-tion networks. As the worldwide pandemic further accelerates the global transformation of optical communication net-works, IOSAC based on silicon photonic technologies are promising to be achieved with versatile functionalities. The IOSAC architecture utilizes the optical frequency channel for biosensing to transmit directly modulated optical signals, thereby the biosensing information can be delivered along with the modulated signal, enabling real-time sensing without affecting the original communication function. 

In this work, we propose and demonstrate a IOSAC system in real time on the CMOS-compatible silicon nitride (SiN) platform. The fabricated IOSAC system exhibits a high sensitivity of 172 nm/RIU and a transmitting ability of 1.25 Gbps OOK optical signal in the NaCl solution, simultaneously. In addition, the IOSAC can be modified as a specif-ic biosensor by surface biofunctionalization, and the point-of-care clinical diagnosis information can be transmitted through the integrated communication capability. IOSAC based on silicon photonic platform shows the potential to real-ize simultaneous high-performance biosensing and optical communication, and further accelerate the development of IoT and 6G networks.

The conceptual illustration of the proposed IOSAC device and the whole system of IOSAC are presented in Fig. 1, in which sensing modules are embedded into well-established fiber network in life communities to achieve S\&C capabilities. In traditional sensing networks, environment information is collected by sensing monitors, and then exchanged by other independent communication networks. Currently, the IOSAC system enables sensing and communication simultaneously. For example, the biological information can be collected at home by IOSAC devices and send to datacenters and hospitals for further data analysis as shown in the Fig. 1. Also, the advanced IOSAC systems can be designed to monitor environ-ment, such as toxic gases, vibration, and temperature. Combined with the ubiquitous fiber network, the collected infor-mation will be delivered significant more efficient, leading to easier city management. As shown in the bottom right of Fig.1, the light with broad spectrum transmits in the well-established optical cables and a part of the light is coupled into the IOSAC terminals. When properties of analyte change with the output of the sensing ring and the received signal, the information about the change of the analyte can be extracted from the signal changes while maintaining the ability to communicate. In addition, optical antennas and on-chip sources can be integrated into IOSAC systems to form fully inte-grated wireless S\&C systems. As shown in the zoomed view of Fig. 1, the low loss strip waveguide and the highly sensitive double-slots waveguide are integrated into a single sensing ring, leading to a relatively high sensitivity of the sensor and a low external loss of the S\&C system. To achieve an optimized IOSAC, numerical studies of the silicon waveguide for simultaneous S\&C are performed. The differences between the strip waveguide and the double-slots waveguide in the aspects of the waveguide structure, electric field distribution, and waveguide sensitivity are analyzed, as shown in the Fig. 2. With the increase of slot waveguide, the equal waveguide sensitivity is increasing.

\section{Structure design and experiments}
\begin{figure}
\centering
\includegraphics[width=0.955\textwidth]{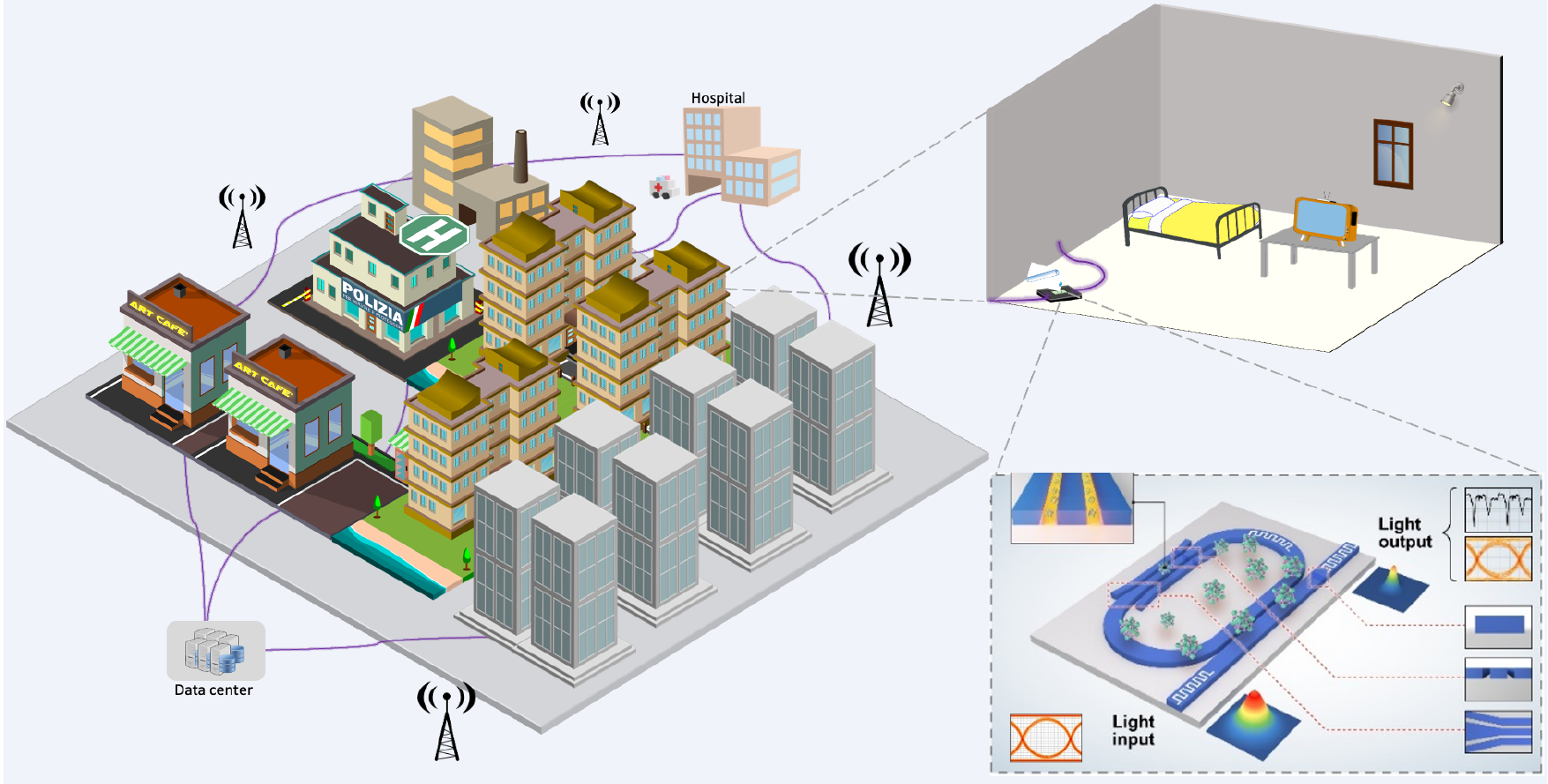}
\caption{\label{fig:Slide1}Conceptual illustration of integrated optical sensing and communication. Enlarge view: the schematic of the proposed ISOAC device.}
\end{figure}

In practice, fiber optic communication systems support a broad spectrum. For example, the entire C-band 35 nm car-ries the signal. To simulate the sensing in real optical communication systems, IOSAC test experiments require a broad-band light source. In the experiments, we use broadband Amplified Spontaneous Emission (ASE) for light amplification and filtering to generate a broadband light source. A sort of optical communication test framework of direct modulation and detection is built up to simulate that the sensing wavelength is added into the optical fiber communication system with a certain bandwidth. Since the microring resonator's resonant wavelength varies depending on the environment, the testing system must have a specified bandwidth so that the change in the sensing wavelength may be noticed within the band. An ASE laser source is deployed, which carries the encoded signal and performs the sensing experiment in the IOSAC testing system. In addition, the spectral characteristics of the all-pass microring present only a few resonant digs and slight energy loss, enabling the realization of the sensing functionality without deteriorating the communication. Thus, the sensor is based on an all-pass microring, and then the S\&C optical signal will be outputted at the through end of the all-pass microring. Finally, the sensing and communication performance of the proposed IOSAC was characterized by analyzing the received S\&C signal, respectively.

\begin{figure}
\centering
\includegraphics[width=0.955\textwidth]{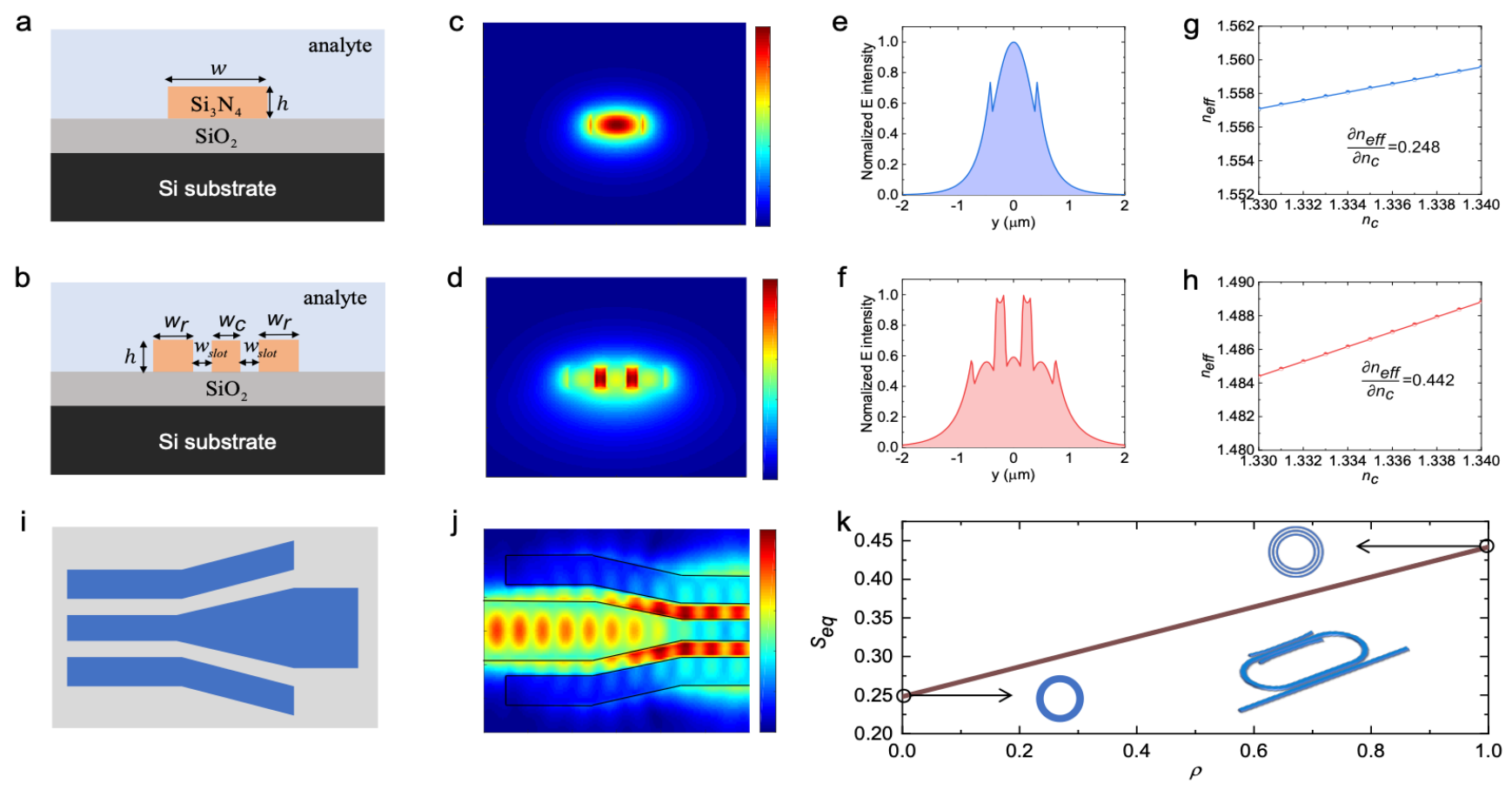}
\caption{\label{fig:Slide2}Schematic cross-section of \textbf{a} the strip waveguide and \textbf{b} the double-slots waveguide; the E-field distribution of \textbf{c} the strip waveguide and \textbf{d} double-slots waveguide; The line plot of the normalized E-field of \textbf{e} the strip waveguide and \textbf{f} the double-slots waveguide; the calculated waveguide sensi-tivity of \textbf{g} the strip waveguide and \textbf{h} the double-slots waveguide; \textbf{i }the top view of the strip-slot mode converter; \textbf{j }the E-field distribution of the strip-slot mode converter; \textbf{k} equivalent waveguide sensitivity $S_{eq}$ as a function of the optimization parameter $\rho$.}
\end{figure}

\section{Results and discussion}

In the experiment, IOSAC devices are fabricated using the standard CMOS process on the SiN platform, as shown in Fig. S1 (Supplementary Materials). The silicon nitride layer thickness is chosen as 400 nm to maintain compact footprint while also boosting low loss. Low-pressure chemical vapor deposition (LPCVD) is typically adopted to fabricate the high-quality SiN films, since the ratio of silicon to nitrogen in the film grown by LPCVD is closer to stoichiometric Si3N4, on which waveguide devices fabricated have the advantages of low optical loss and durability. However, large wafer tensile stress is caused by the LPCVD process, and the stress increases with the increase of the film’s thickness, which may even cause wafer cracking. In this work, a two-step LPCVD method of the SiN film was developed, followed by a high-temperature annealing process to release the stress accumulated in the SiN film. The deep ultraviolet (DUV) lithography inductively coupled plasma (ICP) etching processes were adopted to define the waveguides, the ring resonators, and the grating couplers on the SiN film. Finally, the second annealing was applied to further reduce the accumulated stress in the SiN film. The strip waveguide, hybrid waveguide, and the slot waveguide-based one were fabricated in the same wafer. The SEM images of the fabricated hybrid-waveguide ring resonator are shown in Fig. 3a. The small blocks in Fig. 3a are dummy layers, which are applied to balance the pattern density to reduce the fabrication non-uniformity. And the zoomed images are the coupling region of the bus waveguide and the hybrid-waveguide microring and the strip-slot mode convert-er. A TE polarized light beam was coupled into the chip from a lensed optical fiber to a waveguide via a focusing grating coupler, which is shown in Fig. 3b. The transmission spectra measurements for the hybrid-waveguide resonator for TE polarization with the gap = 400 and 500 nm are shown in Fig. 3c. The measured Q-factors of the proposed hybrid-waveguide resonator with gap = 400 nm and 500 nm are approximately 8000 and 11000 while extinction ratios (ER) of them are larger than 9 dB and 6.5 dB, respectively.

To demonstrate the proposed device with high sensitivity and robust optical signal transmission ability, the sensing and communication experiments are performed. NaCl concentration is used as indicator of biological and environmental concerns for proof of the concept, and thus the S\&C chip is placed in different concentrations of NaCl solution to detect its concentration while transmitting OOK signal. The S\&C experiment setup is shown in Fig. 4. The spontaneous emis-sion light emitted from ASE (ACCELINK ASE-C/L-F-12-FC/APC) is amplified by EDFA (Amonics AEDFA-27-B-FA) and filtered by OTF (Santec, OTF-350), leading to a filtered light with 5 nm band range. Subsequently, the filtered light is adjusted by the PC and then enters the modulator (OM5757C-CTM388) with 1.25 Gbps OOK with a length of 2$^{31}$-1 modulated signals. The initial output power of the modulator is -1.2 dBm and after being amplified by EDFA1 (AEDFA-23-B-FA), the output power rises to 23 dBm. The light is coupled into the SiN S\&C chip via the grating coupler. The light propagates through the S\&C chip with complete signal and sensing information collected from the environment. The output light carrying two types of information is amplified by EDFA2 after transmitting in optical fiber, and then passed through the ATT. The light is divided into two channels through a 50:50 power splitter. One channel is connected to the spectrometer for spectrum measurement, and the other channel is connected to the BERT (T-BertQ J1601A-4) to measure the bit error rate through the PD. It is worth noting that the BER and the spectrum can be observed in real time through-out the experiment. The variation of NaCl concentration is monitored in real time by the shift of the resonance peak in the spectrum, and the communication quality can be observed in real time by the BER. Thus, the sensing and communication functionalities are achieved simultaneously through embedding the SiN sensing resonator into the conventional optical communication systems.

\begin{figure}
\centering
\includegraphics[width=0.955\textwidth]{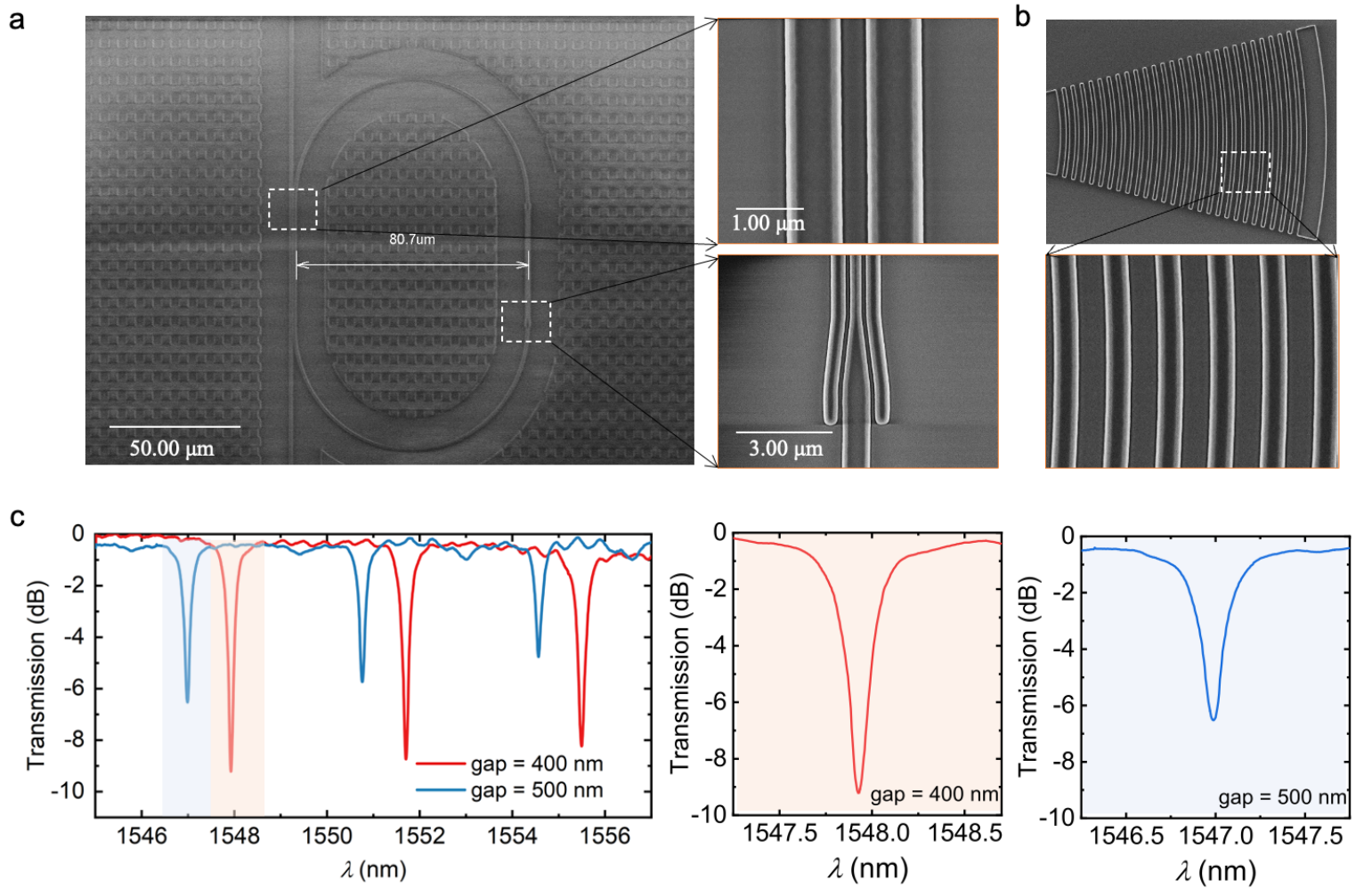}
\caption{\label{fig:Slide3}\textbf{a} The top-view SEM images of the hybrid-waveguide ring resonator with the radius = 40 $\mu m$; The magnification shows the strip-slot mode convert-er and the gap between bus waveguide and ring; \textbf{b} the SEM images of the focusing grating coupler and its magnification; \textbf{c} Measured transmission spectra of the hybrid-waveguide micro-ring resonator for TE polarization with the gap = 400 and 500 $nm$, and zoomed-in view of the resonant dip.}
\end{figure}

\begin{figure}
\centering
\includegraphics[width=0.955\textwidth]{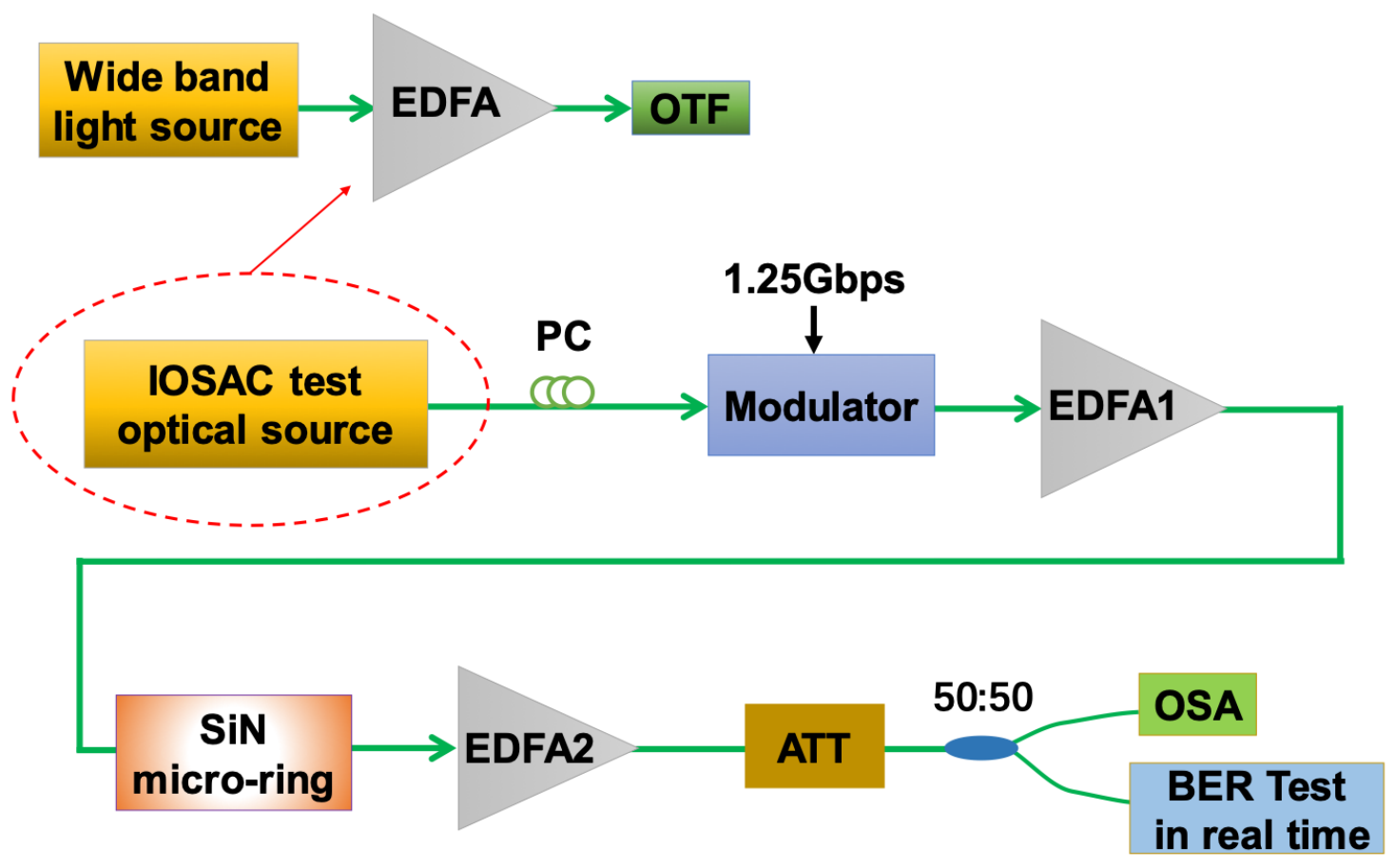}
\caption{\label{fig:Slide6}The IOSAC testing experiment setup. EDFA: Erbium-Doped Fiber Amplifier; OTF: Optical Tunable Filter; ATT: Attenuator; PD: Photo Diode De-tector; OSA: Optical Spectrum Analyzer; BERT: Bit Error Rate Test.}
\end{figure}

Fig. 5a exhibits the shift of the resonant wavelength due to the variation of the concentration of the NaCl solutions. Fig. 5b shows the resonance wavelength shift $\delta\lambda$ versus the refractive index nc, indicating that the sensitivity of the pro-posed IOSAC device is 172 nm/RIU. Moreover, the dependence on thermal sensitivity is characterized and shown in Fig. 5c. It is observed that the resonant wavelength shift with temperature for the proposed device is relatively small. The thermal sensitivity of the hybrid ring resonator is 14.6 pm/℃, which is much smaller than the thermal sensitivity of the Si-based ring resonator (~90 pm/℃). The hybrid-waveguide structure enables high sensitivity of the sensing functionality and low transmission loss of the communication functionality. Compared with the hybrid waveguide structure, the sens-ing performance of the strip waveguide and slot waveguide-based sensing resonator are 97 nm/RIU and 197 nm/RIU, respectively. Though the slot waveguide-based resonator possesses higher sensitivity, it suffers high propagation loss which causes significant signal decay.

The relationship between BER and received power is shown in Fig. 5d. The horizontal axis “received power” is the power received by the power meter, it’s equal to the optical power received by the PD. The optical signal is received by PD, then enters into the bit error meter to measure the bit error rate (BER), which is shown in the longitudinal axis. The BER is measured in five cases, of which back-to-back is measured without the chip, “DI water” is the test result of drip-ping deionized water on the S\&C chip, and the other three cases are NaCl solution of different concentrations (mass ratio): 2.5\%, 5.0\%, and 10.0\%. It can be seen from the results that deionized water influences the bit error rate because water absorbs light. The higher the concentration of NaCl solution, the lower the bit error rate, because the electromagnetic wave attenuates greatly in NaCl solution, and the attenuation increases with the increase of concentration. The experimental results of sensing and optical signal transmission have demonstrated the functions of sensing and communication of the proposed IOSAC chip. The inset of Fig. 5d is the receiver eye diagrams with a BER at the 10$^-6$ level (the operation wave-length is 1546 nm). We change the received power through the optical attenuator to test the “BER vs received power” curve to observe the influence of the S\&C chip on communication performance. In this experiment, the BER can be main-tained at 10$^-7$. Even if the error rate is less than 10$^{-7}$, the error correction can be realized by hard decoding forward error correction (FEC). The SiN micro-ring chip with different sizes are expected to be cascaded along the fiber link to realize a distributed sensor network and detect other parameters such as temperature, strain, and stream velocity. 
In this conceptual S\&C experiment, we utilize a light source with a large spectral width (about 5nm) to observe the sensing information. When using the broadband laser to add the modulating signal, due to the existence of gaps in the spectrum, it will destroy the integrity of the sensing signal once the baud rate of the modulated signal is too high, so the baud rate of the modulated signal is difficult to improve. On the other hand, the focus in this experiment is using the optical sensing signal to transmit low-speed sensing information and communication control information or achieving a low-speed backup optical communication link under extreme conditions, so the 1.25 Gbps OOK can meet the demand basically.

\begin{figure}
\centering
\includegraphics[width=0.9555\textwidth]{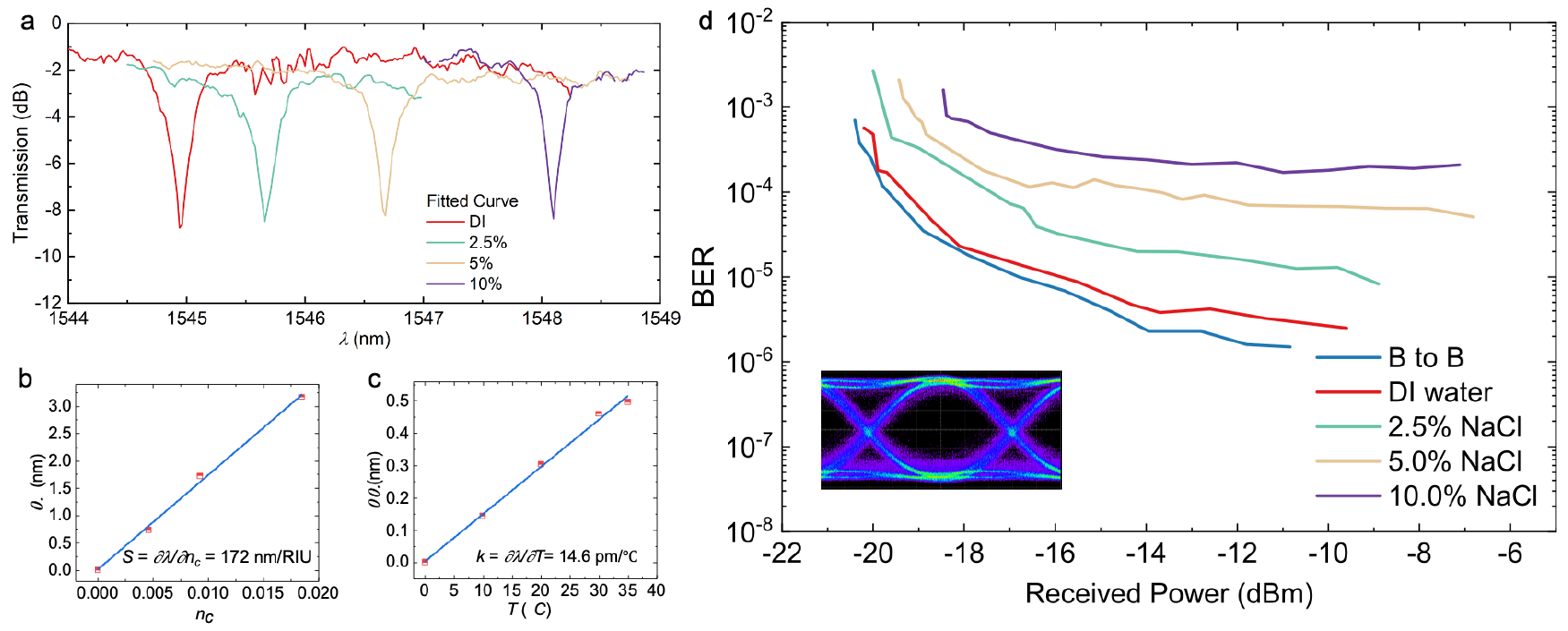}
\caption{\label{fig:Slide4}\textbf{a} Normalized transmission spectra of the hybrid-waveguide ring resonator with four various ambient refractive indices, corresponding to different DI H2O-NaCl mixtures; \textbf{b} resonance wavelength shifts versus refractive index changes;\textbf{ c} resonance wavelength shifts versus temperature chang-es; \textbf{d} BER performance versus received power for B to B, DI water, 2.5\% NaCl, 5.0\% NaCl and 10.0\% NaCl. Inset: eye diagram@BER=10$^{-6}$.}
\end{figure}

The IOSAC technology is a promising research direction for next-generation networks. The input and output of opti-cal signals of the proposed chip are achieved by the coupling between the optical fiber and the grating coupler, while these signals transmission can be realized by the optical antenna based on silicon photonics technology. Therefore, the applica-tion scenarios of IOSAC are extended to the wireless IoT and networks, accelerating the development of the intelligent world. In addition, the demonstration is not limited to silicon photonics in the near-infrared ranges but can be extended to other wavelength range and materials, such as middle to far-infrared wavelengths range and subwavelength plasmonic systems.

\section{Conclusion}

Integrated optical communication and sensing configuration based on a hybrid-waveguide structure on the SiN platform, as well as the concept of IOSAC, have been proposed and demonstrated in this work. Simultaneous S\&C experimental verification system was built, and the experimental results showed that the S\&C chip can simultaneously carry out refrac-tive index sensing and optical signal transmission in an over 10 km long-haul optical fiber system. In the experiment, 1.25 Gbps OOK optical signals were transmitted in 2.5\%, 5\%, and 10.0\% NaCl solution, respectively. The experimental results showed that the higher the NaCl solution concentration, the longer the resonance peak of the micro-ring moves to the long wavelength, and the NaCl solution concentration can be inversely deduced by the shift of the peak wavelength, and the 1.25 Gbps OOK optical signal can be transmitted with high quality in various concentrations of NaCl solution. These results exhibit that sensing and communication functionalities can be integrated into a single system to accomplish IOSAC, leading to reductions in both hardware and signaling costs as well as compact product size and enhanced spectral efficiency for next-generation networks.

\section{Acknowledgments}

 \textbf{Research funding:} This work was funded by the National Natural Science Foundation of China (Grant Number: 62274179, 61904196, 62235001, 62105250, 62275208), National Key R\&D Programme (2022YFB2802400). \textbf{Conflict of interest statement:} The authors declare no conflicts of interest regarding this article.

\section*{References}

\begin{flushleft}
[1]	F. Liu et al., "Integrated Sensing and Communications: Towards Dual-functional Wireless Networks for 6G and Beyond," IEEE Journal on Selected Areas in Communications, pp. 1-1, 2022, doi: 10.1109/JSAC.2022.3156632.

[2]	S. Tian, X. Zhang, X. Wang, J. Han, and L. Li, "Recent advances in metamaterials for simultaneous wireless information and power transmission," Nanophotonics, vol. 11, no. 9, pp. 1697-1723, 2022, doi: 10.1515/nanoph-2021-0657.

[3]	S. Dang, O. Amin, B. Shihada, and M.-S. Alouini, "What should 6G be?," Nature Electronics, vol. 3, no. 1, pp. 20-29, 2020, doi: 10.1038/s41928-019-0355-6.

[4]	P. Zhang et al., "Toward Wisdom-Evolutionary and Primitive-Concise 6G: A New Paradigm of Semantic Communication Networks," Engineering, vol. 8, pp. 60-73, 2022/01/01/ 2022, doi: https://doi.org/10.1016/j.eng.2021.11.003.

[5]	F. Liu et al., "Integrated Sensing and Communications: Toward Dual-Functional Wireless Networks for 6G and Beyond," IEEE Journal on Selected Areas in Communications, vol. 40, no. 6, pp. 1728-1767, 2022, doi: 10.1109/jsac.2022.3156632.

[6]	R. Zhang et al., "Ultracompact and low-power-consumption silicon thermo-optic switch for high-speed data," Nanophotonics, vol. 10, no. 2, pp. 937-945, 2020, doi: 10.1515/nanoph-2020-0496.

[7]	W. Shi, Y. Tian, and A. Gervais, "Scaling capacity of fiber-optic transmission systems via silicon photonics," Nanophotonics, vol. 9, no. 16, pp. 4629-4663, 2020, doi: 10.1515/nanoph-2020-0309.

[8]	S. Bannur Nanjunda et al., "Emerging nanophotonic biosensor technologies for virus detection," Nanophotonics, vol. 11, no. 22, pp. 5041-5059, 2022, doi: 10.1515/nanoph-2022-0571.

[9]	N. Li et al., "Fully integrated electrically driven optical frequency comb at communication wavelength," Nanophotonics, vol. 11, no. 13, pp. 2989-3006, 2022, doi: 10.1515/nanoph-2022-0146.

[10]	V. Cisco, "Cisco visual networking index: Forecast and methodology, 2016–2021," CISCO White paper, 2017.

[11]	P. J. Winzer, D. T. Neilson, and A. R. Chraplyvy, "Fiber-optic transmission and networking: the previous 20 and the next 20 years [Invited]," Opt. Express, vol. 26, no. 18, pp. 24190-24239, 2018/09/03 2018, doi: 10.1364/OE.26.024190.

[12]	B. Corcoran et al., "Ultra-dense optical data transmission over standard fibre with a single chip source," Nature Communications, vol. 11, no. 1, pp. 1-7, 2020.

[13]	E. L. Wooten et al., "A review of lithium niobate modulators for fiber-optic communications systems," IEEE Journal of selected topics in Quantum Electronics, vol. 6, no. 1, pp. 69-82, 2000.

[14]	W. Saad, M. Bennis, and M. Chen, "A Vision of 6G Wireless Systems: Applications, Trends, Technologies, and Open Research Problems," IEEE Network, vol. 34, no. 3, pp. 134-142, 2020, doi: 10.1109/MNET.001.1900287.

[15]	X. Wang, Z. Fei, J. A. Zhang, J. Huang, and J. Yuan, "Constrained Utility Maximization in Dual-Functional Radar-Communication Multi-UAV Networks," IEEE Transactions on Communications, vol. 69, no. 4, pp. 2660-2672, 2021, doi: 10.1109/TCOMM.2020.3044616.

[16]	L. Tombez, E. J. Zhang, J. S. Orcutt, S. Kamlapurkar, and W. M. J. Green, "Methane absorption spectroscopy on a silicon photonic chip," Optica, vol. 4, no. 11, 2017, doi: 10.1364/optica.4.001322.

[17]	P. Dong, Y.-K. Chen, G.-H. Duan, and D. T. Neilson, "Silicon photonic devices and integrated circuits," Nanophotonics, vol. 3, no. 4-5, pp. 215-228, 2014, doi: doi:10.1515/nanoph-2013-0023.

[18]	H. Zhou, Y. Zhao, Y. Wei, F. Li, J. Dong, and X. Zhang, "All-in-one silicon photonic polarization processor," Nanophotonics, vol. 8, no. 12, pp. 2257-2267, 2019, doi: 10.1515/nanoph-2019-0310.

[19]	N. Margalit, C. Xiang, S. M. Bowers, A. Bjorlin, R. Blum, and J. E. Bowers, "Perspective on the future of silicon photonics and electronics," Applied Physics Letters, vol. 118, no. 22, May 31 2021, Art no. 220501, doi: 10.1063/5.0050117.

[20]	A. H. Atabaki et al., "Integrating photonics with silicon nanoelectronics for the next generation of systems on a chip," Nature, vol. 556, no. 7701, pp. 349-354, Apr 2018, doi: 10.1038/s41586-018-0028-z.

[21]	C. Xiang et al., "High-Performance Silicon Photonics Using Heterogeneous Integration," IEEE Journal of Selected Topics in Quantum Electronics, vol. 28, no. 3, pp. 1-15, 2022, doi: 10.1109/JSTQE.2021.3126124.

[22]	A. Rao et al., "Towards integrated photonic interposers for processing octave-spanning microresonator frequency combs," Light: Science \& Applications, vol. 10, no. 1, p. 109, 2021/05/26 2021, doi: 10.1038/s41377-021-00549-y.

[23]	C. Xiang et al., "Laser soliton microcombs heterogeneously integrated on silicon," Science, vol. 373, no. 6550, pp. 99-103, 2021, doi: doi:10.1126/science.abh2076.

[24]	Z. Zhou et al., "Prospects and applications of on-chip lasers," eLight, vol. 3, no. 1, 2023, doi: 10.1186/s43593-022-00027-x.

[25]	E. H. Mordan et al., "Silicon Photonic Microring Resonator Arrays for Mass Concentration Detection of Polymers in Isocratic Separations," Analytical Chemistry, vol. 91, no. 1, pp. 1011-1018, 2019/01/02 2019, doi: 10.1021/acs.analchem.8b04263.

[26]	C. Liu et al., "Silicon/2D-material photodetectors: from near-infrared to mid-infrared," Light: Science \& Applications, vol. 10, no. 1, pp. 1-21, 2021.

[27]	S. Meister et al., "Silicon photonics for 100 Gbit/s intra-data center optical interconnects," in Optical Interconnects Xvi, vol. 9753, H. Schroder and R. T. Chen Eds., (Proceedings of SPIE, 2016.

[28]	C. V. Poulton et al., "Long-Range LiDAR and Free-Space Data Communication With High-Performance Optical Phased Arrays," IEEE Journal of Selected Topics in Quantum Electronics, vol. 25, no. 5, pp. 1-8, 2019, doi: 10.1109/jstqe.2019.2908555.

[29]	A. Dasgupta, M.-M. Mennemanteuil,M. Buret, N. Cazier, G. Colas-des-Francs, and A. Bouhelier, "Optical wireless link between a nanoscale antenna and a transducing rectenna," Nature Communications, vol. 9, May 18 2018, Art no. 1992, doi: 10.1038/s41467-018-04382-7.

[30]	J. Tao et al., "Mass-Manufactured Beam-Steering Metasurfaces for High-Speed Full-Duplex Optical Wireless-Broadcasting Communications," Advanced Materials, vol. 34, no. 6, p. 2106080, 2022/02/01 2022, doi: https://doi.org/10.1002/adma.202106080.

\end {flushleft}

\end{document}